\documentclass{EPS}
\usepackage{fix-cm}
\usepackage{amsmath}
\usepackage{graphicx}
\usepackage{multirow}
\usepackage{makecell}
\usepackage[authoryear]{natbib}
\usepackage{url}
\usepackage[utf8]{inputenc}
\usepackage{amssymb}
\usepackage{array}
\usepackage{kantlipsum}
\usepackage{tabularx}
\usepackage{color}
\usepackage{lineno}
\usepackage{subcaption}
\usepackage{xcolor}
\title{Observing UT1-UTC with VGOS}
\author{
R{\"u}diger~Haas, Department of Space, Earth and Environment,\ 
Chalmers University of Technology, Onsala Space Observatory,
SE-439~92 Onsala, Sweden,\
rudiger.haas@chalmers.se\\
Eskil Varenius, Department of Space, Earth and Environment,\ 
Chalmers University of Technology, Onsala Space Observatory,
SE-439~92 Onsala, Sweden,\
eskil.varenius@chalmers.se\\
Saho Matsumoto, Geospatial Information Authority of Japan,\
1 Kitasato, Tsukuba city,\
305-0811 Japan,\
matsumoto-s96n2@mlit.go.jp\\
Matthias Schartner, Department of Geodesy and Geoinformation,\ TU Wien,
Wiedner Hauptstraße 8-10\
AT-1040 Vienna, Austria,\
(now at Institute of Geodesy and Photogrammetry, ETH Zürich, CH-8093 Zürich, Switzerland, mschartner@ethz.ch)
}
\abstract{
We present first results of UT1-UTC determinations using the VLBI Global Observing System (VGOS).
During December 2019 through February 2020 a series of 1~hour long observing sessions were performed using the VGOS stations at Ishioka in Japan and the Onsala twin telescopes in Sweden.
The data of this VGOS-B series were correlated, post-correlation processed, and analysed at the Onsala Space Observatory.
The derived UT1-UTC results were compared to corresponding results from standard legacy S/X Intensive sessions (INT1/INT2), as well to the final values of the International Earth Rotation and Reference Frame Service (IERS), provided in IERS Bulletin~B.
The VGOS-B series achieve 3-4 times lower formal uncertainties for the UT1-UTC results than standard legacy S/X INT series.
Furthermore, the root mean square (RMS) agreement with respect to the IERS Bulletin~B is 30-40~\% better for the VGOS-B results than for the INT1/INT2 results.  
}

\begin{document}

\maketitle

\section{Introduction}
Geodetic Very Long Baseline Interferometry (VLBI) \citep{Sovers_et_al_1998} is to date the only space geodetic technique that can observe all earth orientation parameters (EOP).
This is primarily due to that VLBI directly observes natural radio sources in the International Celestial Reference Frame \citep[ICRF,][]{Charlot_et_al_2020} with radio telescopes on the surface of the earth, i.e. in the terrestrial reference frame \citep[ITRF,][]{Altamimi_et_al_2016}.
The EOP are the transformation parameters between these two frames.
The EOP that is most difficult to predict due to rapidly varying geophysical excitation is the daily rotation of the earth.
It is usually reported as difference between UT1 and Universal Time Coordinated (UTC), and a precise and accurate monitoring of this parameter is of great importance for satellite navigation systems and satellite orbit determination
\citep{Bradley_et_al_2016}.

The International VLBI Service for Geodesy and Astrometry (IVS) \citep{Nothnagel_et_al_2017} therefore organises dedicated regular observation sessions to determine UT1-UTC, so that the International Earth Rotation and Reference Frames Service (IERS) can produce precise, accurate and reliable data series of UT1-UTC with low latency that scientific users and society at large can use.
These IVS session are the so-called IVS Intensive sessions (INT), which have been observed routinely since decades with the legacy S/X  VLBI system that the IVS organises.
The INT series make use of long east-west oriented baselines due to their high sensitivity for UT1.
The two main series are INT1, usually observed on weekdays on the baseline between Kokee (Hawaii, US) and Wettzell (Germany), and INT2, usually observed on weekends on the baseline between Wettzell (Germany) and Ishioka (Japan).
Before Ishioka was involved in INT2, instead the station Tsukuba (Japan) was part of this series.
There is also a third INT series, INT3, which observes on Monday mornings usually with a three-station network.
During the years, a number of variations were seen in terms of which stations were involved, mainly due to replacing stations for times of station outages due to e.g. maintenance.

While the standard INT are observed with the IVS legacy S/X system, during the last years new stations for the next generation VLBI system called the VLBI Global Observing System (VGOS) have been constructed.
VGOS makes use of very fast slewing telescopes with broadband receiving systems covering four frequency ranges, and dual polarization (H/V) capability \citep{Petrachenko_et_al_2009, Niell_et_al_2018}.

Ishioka \citep{Wakasugi_et_al_2019} is one of these stations.
Furthremore, it can exchange the receiving system, i.e. Ishioka can be used for some months of the year as legacy S/X station and for other months of the year as VGOS station.
Another example of VGOS stations are the Onsala twin telescopes \citep[OTT,][]{Haas_et_al_2019}, which are the currently only  operational VGOS twin telescopes.
Both Ishioka and the OTT are participating routinely to the IVS VGOS operations series (VO).
In discussions with the IVS coordinating center in late 2019, the idea came up to start test observations in order to make use of VGOS for Intensive sessions and thus explicitly UT1-UTC determination.
Using the Onsala twin telescopes in this so-called VGOS-B series allows simultaneous UT1-UTC determination with two parallel long east-west baselines connecting to Ishioka.

\section{Scheduling and observing the VGOS-B sessions}
For the period December 2019 through February 2020 in total 12 VGOS-B sessions were scheduled.
The scheduling was done with the software VieSched++ \citep{Schartner_Boehm_2019} involving the VGOS stations ISHIOKA (Is) in Japan and ONSA13NE (Oe) and ONSA13SW (Ow) in Sweden.
The schedules were prepared to be simultaneous to standard INT1 observations.
During scheduling, special emphasis was given to include observations at the corners of the mutually visible sky since these observations are known to provide the most impact on the precision of the derived UT1-UTC results \citep{Uunila_et_al_2012, Gipson_Baver_2015}.
To~achieve this, a special scheduling algorithm was used. 
The schedule starts by observing a source at one of the two corners of the mutually visible sky, followed by scans selected using standard geodetic scheduling optimization. 
Every ten minutes, the algorithm forces a scan of a source located in one of the corners, alternating between the two possibilities. 
Therefore, it is ensured that observations at the corners of the mutually visible sky are well represented in the schedule.

Due to the modern and fast-slewing VGOS telescopes, the number of scheduled observations per baseline was more than 50 during the 1~hour long sessions.
This is significantly more than the usually 20-25 observations during 1~hour long INT1 and INT2 sessions with legacy S/X stations.
We note that the VGOS-B sessions analysed in this paper were scheduled assuming standard VO-session recording overheads, such as buffer-flush times needed for Mark6-recording systems, which limit the number of scans per unit time. 
Future VGOS-B sessions will be further optimised for the Is-Oe/Ow systems, which do not have these overheads, and can therefore schedule $\sim100$ observations per baseline per hour, fully utilising the fast slewing speed.

Table~\ref{TAB:VGOS-B} lists these 12 VGOS-B sessions and their INT1 counter parts, including the stations involved.
From simulations with VieSched++ the formal uncertainties of the UT1-UTC estimates are 4-5~$\mu$s for the VGOS-B sessions, while they are a factor of 2-3 worse for the legacy S/X intensive schedules.
Unfortunately, during four of the twelve VGOS-B sessions technical problems occurred at Ishioka, so that the observed raw data were not complete, see the comments in the table. 

\begin{table*}[ht!]
\centering
\caption{VGOS-B sessions in December 2019 through February 2020, and their simultaneously observed INT1 counterparts.
The stations are ISHIOKA (Is), ONSA13NE (Oe), ONSA13SW (Ow), KOKEE (Kk), WETTZELL (Wz), SVETLOE (Sv) and MK-VLBA (Mk).}
\begin{tabular}{ l l  l | l l }
\hline
Code & Stations & Comment & Code & Stations \\
\hline
B19344 & Is-Oe-Ow &                                 & I19344 & Mk-Wz \\
B19351 & Is-Oe-Ow &                                 & I19351 & Kk-Wz \\
B19357 & Is-Oe-Ow &                                 & I19357 & Kk-Wz \\
B19364 & Is-Oe-Ow &                                 & I19364 & Kk-Wz \\
B20007 & Is-Oe-Ow & (Is-data: only VGOS band-A)     & I20007 & Kk-Wz \\ 
B20013 & Is-Oe-Ow &                                 & I20013 & Kk-Wz \\
B20023 & Is-Oe-Ow &                                 & I20023 & Kk-Wz-Sv \\
B20027 & Is-Oe-Ow &                                 & I20027 & Kk-Wz \\
B20037 & Is-Oe-Ow &                                 & I20037 & Kk-Wz-Sv \\
B20044 & Is-Oe-Ow & (Is-data: only H-polarization)  & I20044 & Kk-Wz \\
B20048 & Is-Oe-Ow & (Is-data: only H-polarization)  & I20048 & Mk-Wz \\
B20055 & Is-Oe-Ow & (Is-data: only H-polarization)  & I20055 & Kk-Wz \\
\hline
\end{tabular}
\label{TAB:VGOS-B}
\end{table*}

\section{Correlation and post-correlation analysis}
The observed raw data of the Ishioka station were transferred electronically to the VLBI correlator at the Onsala Space Observatory.
The software correlator DiFX version 2.6.1 \cite{Deller_et_al_2011} was used for correlation and the Haystack Observatory Postprocessing System \citep[HOPS,][]{HOPS} version 3.21 was used for the post-correlation analysis. 
The data were processed according to the VGOS Data Processing Manual \citep{VGOSproc}. This included forming the pseudo-stokes I polarisation product, from the recorded H and V data, to account for parallactic-angle differences. Both H and V are required for this process, and therefore the single-polarisation Is-data (see Table \ref{TAB:VGOS-B}) were omitted from post-processing.
The short (75~m) baseline Oe-Ow suffered from disturbing local radio frequency interference (RFI). The full VGOS band-A, and a few channels in other VGOS-bands, were therefore exluded from Oe-Ow post-processing, thus deteriorating the delay accuracy on this baseline.
Finally, the VLBI delay observations were exported to vgosDb format \citep{Bolotin_et_al_2015}.
The corresponding tools of the $\nu$Solve software package version 0.7.1 \citep{Bolotin_et_al_2012} were used to calculate and add the theoretical delays (vgosDbCalc) as well as the supplemental data from the observing station logfiles (vgosDbProcLogs). 
We note that while Oe and Ow are equipped with Cable Delay Measurement Systems (CDMS) which stores delay corrections in the observing logfiles, Is currently does not measure such corrections. 
Therefore, proxy-cable corrections were derived using the Is phase-calibration data as described in VGOS Data Processing Manual \citep{VGOSproc}.

\section{Data analysis and results}
The geodetic data analysis was performed with the ASCoT software \citep{Artz_et_al_2016}.
We analysed both the VGOS-B sessions, as well as all INT1 and INT2 sessions in the time range December 2019 through February 2020.
As mentioned above, only eight of the twelve VGOS-B sessions could be analysed, since during four of these sessions some technical problems had occurred at Ishioka station.

A~standard analysis approach for INT sessions was used:
\begin{itemize}
\item All station coordinates were kept fixed on their a~priori values, i.e. the IVS VTRF2019d \citep{VTRF2019d} values.
For the OTT we used the corresponding VTRF2019d coordinates that were determined through dedicated short-baseline interferometry campaigns \citep{Varenius_Haas_2020}.
\item The radio source positions were kept fixed on their ICRF3 \citep{Charlot_et_al_2020} values.
\item A~priori information for the EOP was taken from {\em \url{usno_finals.erp}} created 2020.08.19-23:00:17.
\item We fixed the clock for one of the stations as reference, while estimating 2nd order clock polynomials for the other stations involved.
\item We estimated one zenith wet delay parameter for each of the stations but atmospheric gradients were not estimated.
\item The UT1-UTC parameter was estimated.
\end{itemize}

\begin{figure*}[htbp!]
\centering
\includegraphics[width=0.95\textwidth]{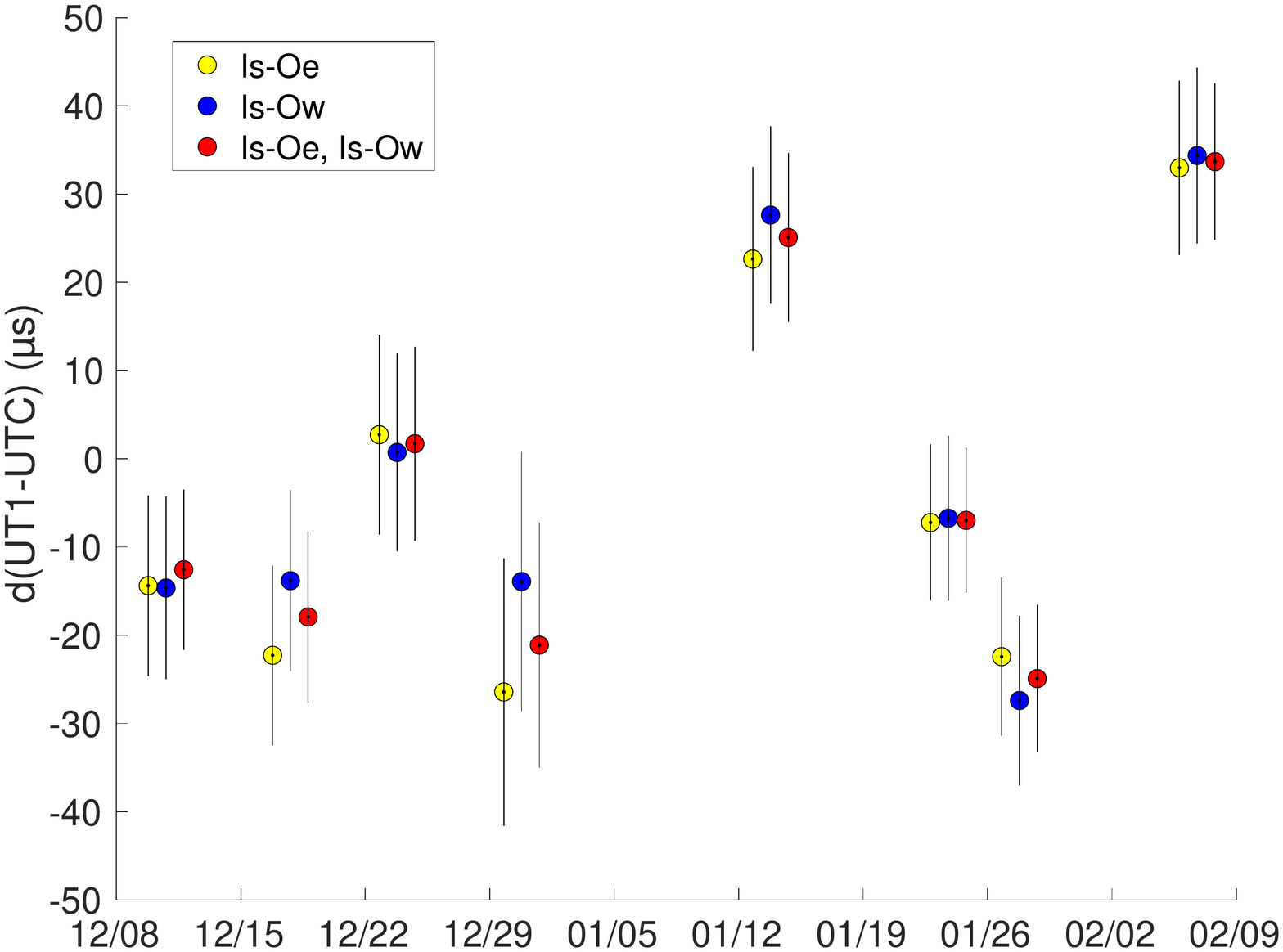}
\caption{Comparsion of UT1-UTC corrections to the a~priori values derived from analysing the VGOS-B baselines individually, i.e. Is-Oe (yellow) and Is-Ow (blue), and in a common analysis (red). 
The uncertainties shown are $1 \sigma$.
}
\label{FIG:VB_UT1}      
\end{figure*}

First, we investigated the agreement of the results of the two parallel baselines Is-Oe and Is-Ow by analyzing the two baselines individually and together.
The short baseline Oe-Ow was excluded from any analysis due to the previously mentioned RFI problems.
Figure~\ref{FIG:VB_UT1} depicts the corrections to the a~priori UT1-UTC from these three approaches.
The results of all the analysis of the two individual baselines, i.e. Is-Oe and Is-Ow, agree within their formal errors.
The analysis of both baselines together provides an average of the two result of the two individual baselines. 
With 4~$\mu$s it also gives the lowest median formal uncertainty of UT1-UTC, while the median formal uncertainties for the individual baselines are 6~$\mu$s and 5~$\mu$s for Is-Oe and Is-Ow, respectively. 
These formal uncertainties correspond well with the results from simulations performed with VieSched++.
In the following we used the results of analysing both baselines together.

Besides performing our own data analysis of the INT1 and INT2 sessions, we also downloaded the results of the corresponding analysis that were provided by five of the IVS Analysis Centres (IVS ACs).
These were the Bundesamt für Kartographie und Geodäsie (BKG), the NASA Goddard Space Flight Center (GSF), the United States Naval Observatory (USN), the Geospatial Information Authority of Japan (GSI), and the Institute of Applied Astronomy of the Russian Academy of Sciences (IAA).
These IVS ACs routinely analyze the INT sessions and provide their results to the IERS for the determination of IERS rapid and final products \citep{IERS_Bull_B}, IERS Bulletin~A and IERS Bulletin~B, respectively.
All except GSI provide results for INT1 and INT2 sessions.
The (few) results with formal uncertainties larger than 100~$\mu$s were excluded from further comparisons.

\begin{figure*}[htbp!]
\centering
\includegraphics[width=0.95\textwidth]{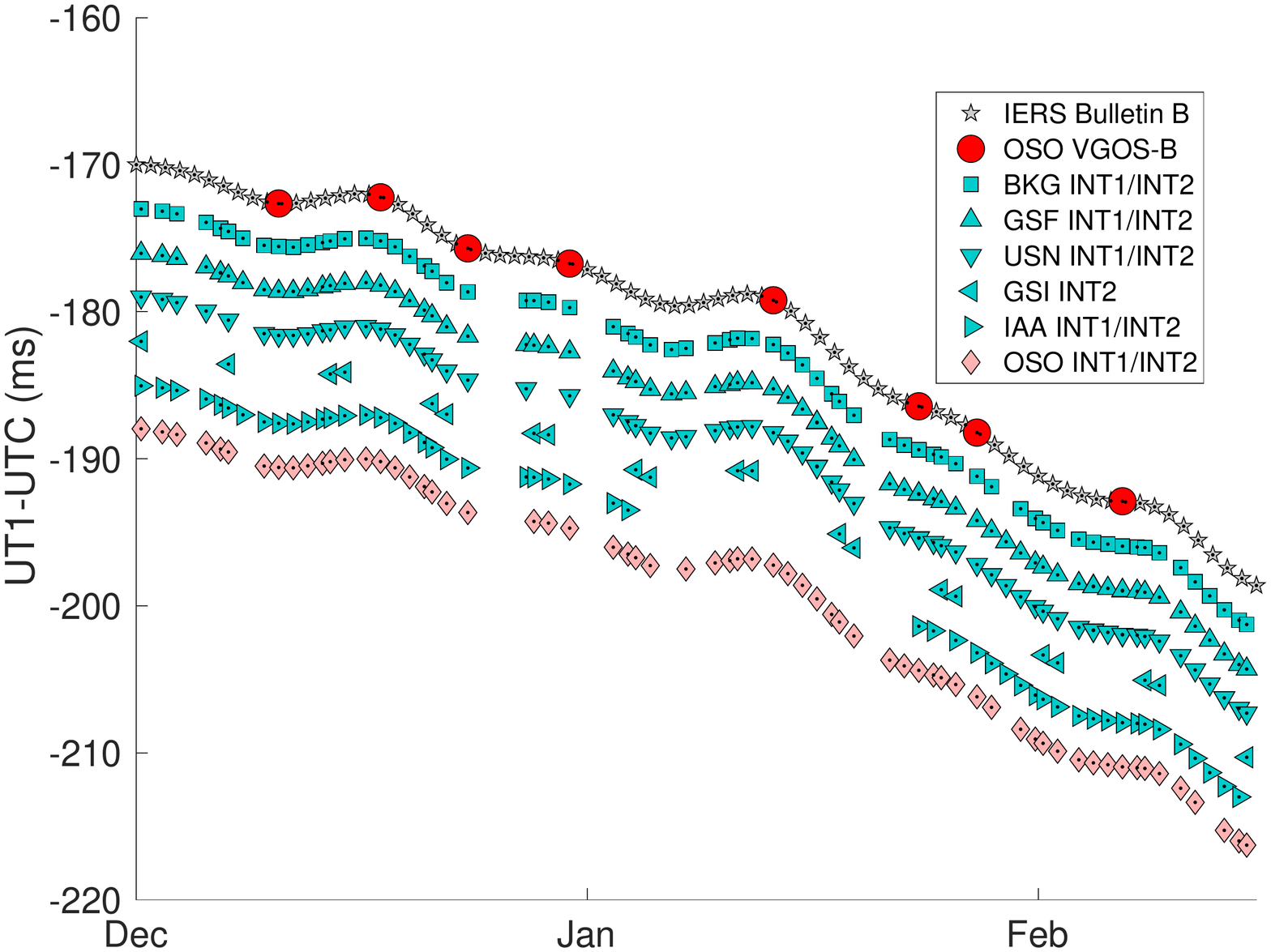}
\caption{Time series of UT1-UTC for December 2019 through mid February 2020.
Shown are IERS Bulletin~B values (grey stars) and the VGOS-B results as (red circles).
The light blue symbols show the INT1/INT2 results of BKG (squares), GSF (upward-pointing triangles),  USN (downward-pointing triangles),
GSI (left-pointing triangles), and IAA (right-pointing triangles).
The INT1/INT2 results from OSO are shown as light-red diamonds.
To increase readability, all series, except the VGOS-B one, are offset by $-3$~ms each from the series shown directly above.
}
\label{FIG:UT1}      
\end{figure*}

Figure~\ref{FIG:UT1} presents the UT1-UTC times series. 
The IERS Bulletin~B values are shown as grey stars and the VGOS-B results as red circles.
The INT1/INT2 results of the five IVS ACs are depicted in light blue as squares (BKG), upward-pointing triangles (GSF), downward-pointing triangles (USN), left-pointing triangles (GSI), and right-pointing triangles (IAA).
Our own analysis results of the INT1/INT2 data are presented as light red diamonds.
To increase the readability of this graph, all time series, except the VGOS-B one, are offset by $-3$~ms each from the series shown directly above.
The formal uncertainties of the presented results are on the order of a few $\mu$s to several tens of $\mu$s and thus not visible in this scale since UT1-UTC is changing a lot during the presented time interval.
However, from a first glance at the graph it is obvious that all presented results show a similar signature and agree rather well.
The VGOS-B results (red circles) do not show any obvious deviation with respect to the other data sets.

\begin{figure}[htbp!]
\centering
\includegraphics[width=0.95\linewidth]{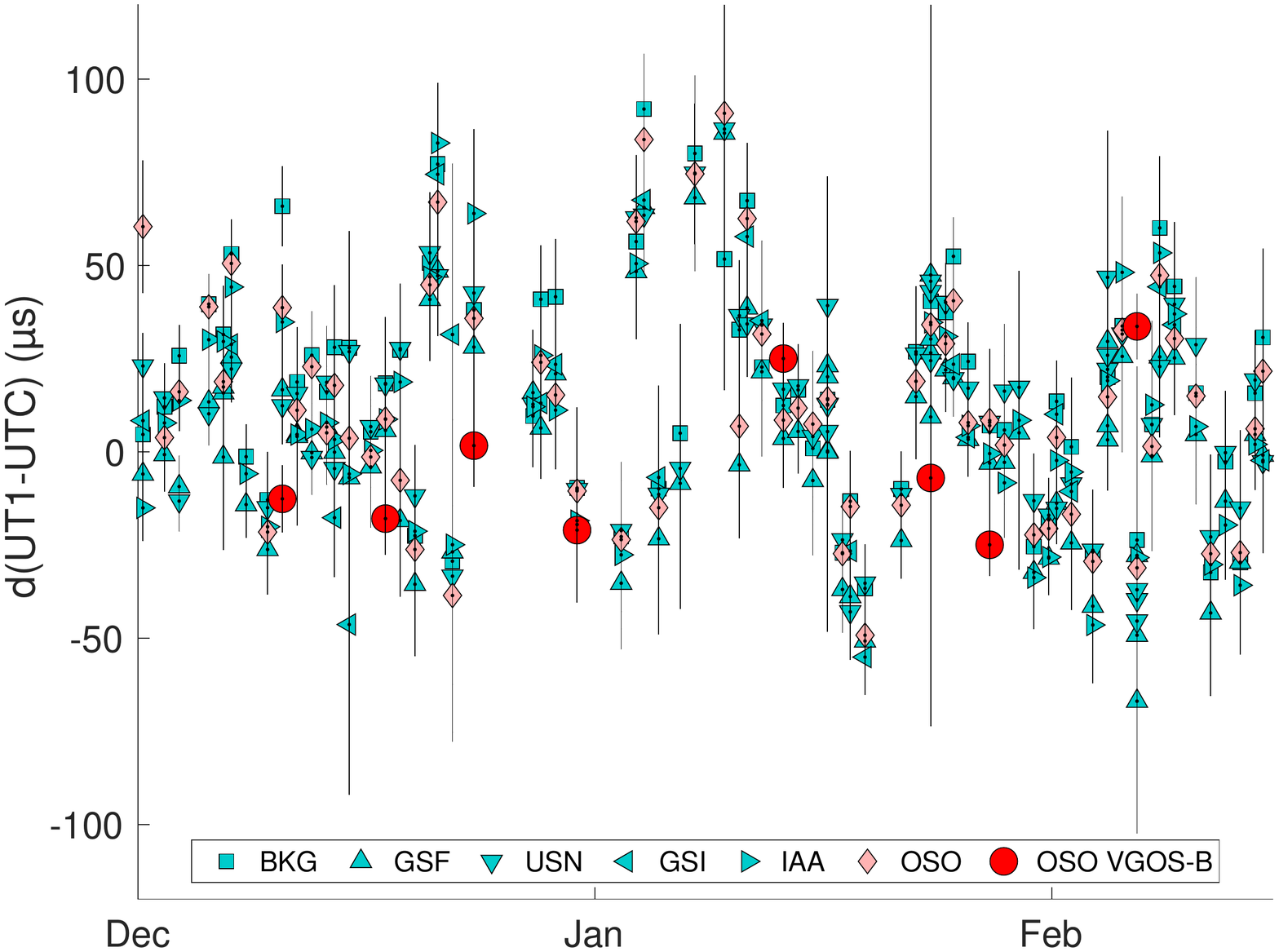}
\caption{UT1-UTC differences w.r.t. to IERS Bulletin~B for December 2019 through mid February 2020.}
\label{FIG:dUT1}
\end{figure}

More insight is gained when investigating the differences between the individual UT1-UTC results and the IERS Bulletin~B values.
The latter are provided as daily values at epoch 00:00 UTC, while the INT1/INT2 and VGOS-B results are given at the respective observation times of the corresponding session.
Thus, to evaluate the agreement between the individual UT1-UTC series and IERS Bulletin~B, the latter had to be interpolated to the epochs of the UT1-UTC series.
For simplicity, and since all of the series do not include any high-frequent variations of UT1-UTC, we used a linear interpolation.
Figure~\ref{FIG:dUT1} depicts the times series of the resulting differences.
It becomes clear that all series agree reasonably well with the IERS Bulletin~B values, with maximum deviations well within the $\pm 100$~$\mu$s interval.
Table~\ref{TAB:statistic} provides both the mean and median formal uncertainties of the individual series, their root mean square (RMS) agreement w.r.t. the IERS Bulletin~B series, the biases, and the remaining scatter after removing the biases, expressed as standard deviation (STD).
The VGOS-B data show both the smallest uncertainties, a small bias, as well as the best agreement in terms of RMS and STD.
The improvements of the VGOS-B results w.r.t. to legacy S/X are between 30~\% and 40~\%, depending on which of the series compared with.

\begin{table*}[htbp!]
\centering
\caption{Statistics on the seven UT1-UTC times series. 
Shown are the mean formal uncertainties ($\sigma_{\rm mean}$), median formal uncertainties ($\sigma_{\rm median}$), the (unweighted) root-mean square agreement (RMS) w.r.t. the IERS Bulletin~B series,
the mean bias and the standard deviation (STD) after removing the bias.
All values are given in unit $\mu$s.
The relatively large uncertainty of the VGOS-B bias reflects the relatively few data points in this series.}
\begin{tabular}{ l | r r r r r }
\hline
Series & $\sigma_{\rm mean}$ & $\sigma_{\rm median}$ & RMS  & bias & STD\\
\hline
BKG INT1/INT2 &  14 & 14 & 34 & $18 \pm 4$ & 30\\
GSF INT1/INT2 &  16 & 13 & 28 &  $2 \pm 3$ & 28\\
USN INT1/INT2 &  16 & 13 & 30 & $12 \pm 3$ & 28\\
GSI INT2      &  13 &  9 & 36 & $14 \pm 7$ & 33\\
IAA INT1/INT2 &  14 & 13 & 29 &  $8 \pm 4$ & 28\\
OSO INT1/INT2 &  16 & 15 & 33 & $13 \pm 4$ & 31\\
OSO VGOS-B    &   4 &  4 & 20 & $-3 \pm 8$ & 20\\
\hline
\end{tabular}
\label{TAB:statistic}
\end{table*}

\section{Conclusions and outlook}

We scheduled, observed, correlated, post-correlation processed and analysed a number of VGOS sessions for the determination of UT1-UTC.
These are the first sessions involving VGOS twin telescopes for this purpose.
The involved VGOS stations were Ishioka (Is) in Japan and the Onsala twin telescopes (Oe, Ow) in Sweden, forming two parallel long east-west oriented baselines, i.e. Is-Oe and Is-Ow.
We compared the derived UT1-UTC values from these VGOS-B sessions with corresponding results from our own analysis of INT1/INT2 data, the corresponding INT1/INT2 results of five other IVS ACs, and the final values of the IERS Bulletin~B.
The mean and median formal uncertainties of the VGOS-B derived UT1-UTC results are three to four times lower than the corresponding formal uncertainties of the INT1/INT2 results.
This is in good agreement with simulations performed based on the actual schedules.
The smallest formal uncertainties are achieved when analysing the two parallel baselines Is-Oe and Is-Ow together.
The RMS agreement w.r.t. to IERS Bulletin~B is better for the VGOS-B results than for the legacy S/X INT1/INT2 results and the VGOS-B results have a small bias only with the smallest remaining standard deviation.
The RMS and STD improvements are on the order of 30-40~\%.  
This improvement confirms the simulation study of \citet{Corbin_Haas_2019} where various scenarios of VGOS observations of UT1-UTC were investigated.
Investigating six months of intensive observations by simulations, \citet{Corbin_Haas_2019} found an improvement of the UT1-UTC accuracy of more than 40~\% when simulating VGOS intensive sessions including one VGOS twin telescope station, compared to standard S/X legacy intensives.
The good agreement of the VGOS-B results with IERS Bulletin~B is also to some extent remarkable since the INT1/INT2 results provided by the operational IVS ACs are input data for the determination of the IERS Bulletin~B, while the VGOS-B results are not.
However, more VGOS-B sessions are necessary in order to verify this good agreement.
The plan is thus to continue the VGOS-B sessions, starting in late November 2020, as soon as the Ishioka station is ready to observe VGOS again. 
With even more ($\sim100$) observations per baseline per hour, we expect these sessions to further improve the UT1-UTC determination. 
Finally, we note the possibility of Oe and Ow participating simultaneously in two different UT1-UTC sessions, a unique capability of the twin telescopes sites which can hopefully be investigated for upcoming VGOS intensive sessions.

\section*{Availability of data and materials}
\noindent
The vgosDbs analysed in this study are publically available via the IVS webpages.
The IERS Bulletin~B results are available via the IERS webpages.
The individual UT1-UTC series of the IVS ACS are available from the IVS.

\section*{Competing interests}
\noindent
The authors declare that there are no competing interests.

\section*{Funding}
\noindent

\section*{Authors' contributions}
\noindent
RH, EV, SM and MS organised VGOS-B series, in coordination with the IVS Coordinating Center.
MS scheduled the VGOS-B experiments.
EV performed the VLBI data correlation and post-correlation analysis and created the final vgosDb.
RH analysed the vgosDB with ASCoT and did the comparison analyses.
All authors read and approved the final manuscript.

\section*{Authors’ information}

\section*{Acknowledgements}
We acknowledge advice given by Mike Titus, Simone Bernhart, John Barret and Sergei Bolotin regarding correlation and post-processing of the VGOS B-sessions.

\bibliographystyle{spbasic}      
\bibliography{References}

\end{document}